\documentclass{elsart}
\usepackage{epsfig}

\begin{document}

\begin{frontmatter}

%\draft

\title{Improved predictions of the standard model and the detection of new 
physics in neutron beta decay.}

\author{A. Garc\'{\i}a$^{\ast}$, and J.L. Garc\'\i a--Luna\thanksref{jlg}$^{\ast ,\dagger}$}
\address{$^{\ast}$Departamento de F\'{\i}sica, Centro de Investigaci\'on y de
Estudios Avanzados del IPN, Apartado Postal 14-740, 07000 M\'exico,
Distrito Federal, M\'exico.\\
$^{\dagger}$Departamento de F\'{\i}sica,  Centro Universitario de Ciencias 
Exactas e Ingenier\'{\i}as, Universidad de Guadalajara, Blvd. Marcelino Garcia 
Barragan 1508, C.P. 44840, Guadalajara Jalisco, M\'exico.}
\date{\today}
\thanks [jlg] {E--mail: jlgarcia@fis.cinvestav.mx}
\maketitle

\begin{abstract}
We improve the current predictions of the standard  model for 
neutron beta decay observables and compare them with the currently available 
ones. Next we study their implications in the possible detection of new 
physics. We discuss where the limitations are and where further efforts 
should be directed.
\end{abstract}

\end{frontmatter}

%\twocolumn
{\bf \em Introduction.} As precision measurements in neutron beta decay 
(n$\beta$d) have become available, interest in detecting physics beyond the
standard model (SM) in this decay has increased steadily\cite{abe2}. However, 
the predictions of the SM for $n\beta d$
observables seem to be severely afflicted by our current inability to calculate 
the leading form factors ratio {\large $\lambda$}$ = g_1/f_1$ and the 
Cabibbo-Kobayashi-Maskawa (CKM) matrix element $V_{ud}$. It imposes on them 
two general restrictions, namely, the $V-A$ structure of the weak vertex and the 
unitary of the CKM 
matrix , respectively. Otherwise these two quantities remain as free 
parameters to be determined from experiment. With current experimental 
error bars a region in the ($\lambda$, R)- plane (R is the decay rate) can be 
determined where the SM remains valid at, say, 90\% C.L. We shall refer to 
this region as the standard model region (SMR). Our main purpose in this 
paper is to reduce this region as much as possible, thereby, improving the 
predictions of the SM. Next we shall study the 
usefulness of this improved region in either detecting new physics or 
imposing more strict bounds on its existence. Finally, we shall discuss 
how future efforts in this respect should be directed.
Another important aspect to discuss is the limitations for the detection 
of new physics in $n\beta d$. As precision measurements improve, the limitations 
will be come significant enough as to render the detection of new physics 
in this decay very obscure or even hopeless.\\
{\bf \em Procedure.} As we just mentioned agreement of the SM with n$\beta$d can 
occur only in the
region of the ({\large $\lambda$}, $V_{ud}$) plane or equivalently of the 
({\large $\lambda$}, $R$) plane where both the $V-A$ and the unitarity
restrictions are satisfied, namely the SMR. Disagreement, and along with it 
the possibility to detect new
physics, corresponds to the region in such planes where
the restrictions are no longer satisfied. We shall refer to this other region 
as the new physics region (NPR). we shall proceed determine these two
regions.\\
One can show that up to a precision of $10^{-4}$ the observables in 
$n\beta d$ depend only on two
parameters, namely, the (CKM) matrix element $V_{ud}$
and the ratio $\lambda$ \cite{mio}. All other contributions up to this precision 
are well determined (through the use of the conserved vector-current hypothesis,
model-independent and reliable estimates of model-dependent radiative
corrections, etc.). We shall use the angular asymmetry coeficients $\alpha_e$, 
$\alpha_{e\nu}$, and $\alpha_\nu$ (imposing the $V-A$ restriction through 
{\large $\lambda$}), the CKM $V_{us}$ and $V_{ub}$ matrix elements (imposing the 
unitarity restriction $V_{ub}= \sqrt{1-V_{ud}^2-V_{us}^2}$), and $R$. Then 
we form a
$\chi^2$ function with six terms. The first four terms compare the theoretical
expressions of the $\alpha_e$, $\alpha_{e\nu}$, $\alpha_\nu$ and $R$ as
functions
of {\large $\lambda$} and $V_{ud}$ with their experimental counterparts, the
fifth term compares $V_{ub}$ of the unitarity square root with its experimental
counterpart, and the last term constrains $V_{us}$ to its experimental value.\\
The precision measurements in n$\beta$d occur in $R$ and $\alpha_e$. The 
state-of-the-art will allow that their error bars be reduced to about one tenth
of their current values \cite{3}. In order to determine the SMR we shall
concentrate on these two quantities. We shall keep the other four fixed at their
present experimental values \cite{pdg}, namely, $\alpha_{e\nu}= -0.0766(36)$, 
$\alpha_\nu = 0.9830(40)$, $V_{ub} = 0.0036(10)$ and $V_{us}=0.2196(23)$. We
have folded the theoretical error into the current value of $R$ to give
$R=1.12879(110)$ \cite{mio}. It is customary to quote the experimental value of 
{\large $\lambda$} instead of $\alpha_e$ and we shall follow this practice, i.e.,
in $\chi^2$ we shall replace $\alpha_e$ by {\large $\lambda$}. There are at 
present \cite{liaud,bopp,yero,rei} four precise 
measurements of {\large $\lambda$} (all them through $\alpha_e$), namely, 
$\lambda_L=1.2660(40)$, $\lambda_B=1.2620(50)$, $\lambda_Y=1.2594(38)$, and 
$\lambda_R=1.2735(21)$. This last value is slightly revised in \cite{abe2}. The
first three are statistically compatible and their average is
$\lambda_{LYB}=1.2624(24)$. One may quote the average of the four,
$\lambda_{LYBR}=1.2687(16)$. However, this average is not statistically
consistent and one should really keep $\lambda_R$ separate from the
other three. At any rate the error bars of $\lambda_{LYB}$ and $\lambda_R$ are
comparable, so one may take a typical $\sigma_\lambda=0.0024$. Because of this
situation we shall cover $5\sigma_\lambda$ to the right and $3\sigma$ to the
left of $\lambda_{LYB}$.\\
For definiteness, we shall work in the ({\large $\lambda$}, $R$) plane. Our
procedure to determine the SMR is to calculate $\chi^2$ at points 
({\large $\lambda$}, $R$) covering the rectangle in this plane with sides
(1.256, 1.274) for {\large $\lambda$} and (1.12549, 1.13209) for $R$. This latter
is the $\pm 3\sigma_R$ range. The values of {\large $\lambda$} and $R$ in the
points ({\large $\lambda$}, $R$) are used as the ``experimental'' central values
in the $\chi^2$ function. For the ``experimental'' error bars we shall consider
three possibilities. First we take $\sigma_\lambda$ and $\sigma_R$ at their
current values, next we take for them one tenth of those current values, and last
we take for them one thousandth of such values.\\
{\bf \em Results.} Our numerical results are displayed in Table 1. There we show
the $90\%$ and $95\%$ ranges for {\large $\lambda$} at sample values of $R$
(which go in steps of one current $\sigma_R$ up to $\pm 3\sigma_R$ around its
present experimental central value). For each value of $R$ we give three
entries; the upper, middle, and lower ones correspond to current error bars,
future one tenth error bars, and, so to speak, ``ideal'' one thousandth error
bars, respectively. All these numbers are easily visualized in Figs. 1-3. The
bands within the dotted lines correspond to the $90\%$ SMR in the three cases
considered. In each of these figures we have also depicted with I, II, and III
the $90\%$CL ellipses around $\lambda_{LYB}$, $\lambda_R$ and $\lambda_{LYBR}$,
respectively. In figures 2 and 3 we also show the $90\%$CL ellipses 
corresponding to one tenth of $\sigma_\lambda$ and $\sigma_R$ around these three central values
and also around a central value previously reported in \cite{2}, $\lambda_{W}=
1.2657(30)$. These figures make it very easy to reach several conclusions. 
The $\lambda_{LYB}$ and $\lambda_R$ with their current error bars both fall
completely at a $90\%$CL in the NPR, i.e., each one gives a strong
signal for the existence of new physics. This cannot be seen in Fig. 1, but it
becomes very clear in Figs. 2 and 3. Of course, since they fall on opposite sides
of SMR, the situation is yet undecided. Another conclusion is that the SMR is
greatly improved by reducing $\sigma_R$ and $\sigma_\lambda$ by one tenth; but,
even though there is some improvement, to further reduce $\sigma_R$ and 
$\sigma_\lambda$ does not provide a substantial gain. The reason is that in Fig.
2 The error bars of $V_{us}$ already dominate. This is clearly seen in Fig. 3,
where the width of the SMR is due almost entirely to the $V_{us}$ error bars.\\
{\bf \em The the detection of new physics in \mbox{\boldmath $n\beta d$}.} 
The ability of n$\beta$d 
to resolve new physics depends on where in the ({\large $\lambda$}, $R$) 
plane the future ({\large $\lambda_{exp.}$}, $R_{exp.}$) will lie as well as on 
their future accompanying error bars. One may appreciate this by looking 
at the values of $\chi^2$ at points covering the rectangle we are studying. For 
this purpose we have produced Table 2. In each entry we give two numbers. The 
upper one uses the current $\sigma_R$ and $\sigma_\lambda$, the lower one uses 
one tenth of $\sigma_R$ and $\sigma_\lambda$. $\chi^2$ does not further
increases
substantially if one uses one thousandth of $\sigma_R$ and $\sigma_\lambda$, but
we shall not produce the corresponding numbers in Table 2. $\chi^2$ may reach
impressively high values.\\
There is another way to appreciate this resolving power. This can be seen in
Table 3. We selected the four values of {\large $\lambda$} marked as I-IV in
Figs. 2 and 3. The entries in this table contain four values of $\chi^2$, which
up-downwards are in the order I-IV. The error bars in {\large $\lambda$} and $R$
are reduced from their present values to one tenth of them in steps $1/N$,
$N=1,2,...,10$. Each one of the {\large $\lambda$} values corresponds to new
physics, as is indicated by the small ellipses in Figs. 2 and 3. In Table 3 we
can see how the resolution increases as the error bars in {\large $\lambda$} and
$R$ decreased.\\
To clearly resolve I and II it is quite sufficient to cut $\sigma_R$ and 
$\sigma_\lambda$ simultaneously to a little over one-half their current values.
To resolve point IV requires that $\sigma_R$ and $\sigma_\lambda$ be 
simultaneously cut to around one sixth. To resolve the overall 
average of {\large $\lambda$} at III
requires more effort , because of its closeness  to the SMR. However, Fig. 3
shows it falls in the NPR. Also, one can see in this table that reducing 
$\sigma_R$
alone is of little use in detecting new physics. While reducing $\sigma_\lambda$
alone helps in cases I and II, but is useless in cases III and IV.\\
{\bf \em Summary and discussion.}  In this paper we have substantially improved the 
predictions of the SM for $n\beta d$ observables, as can be appreciated by 
comparing 
Fig. 1 with Figs. 2 and 3. Our analysis has several implications. First, it 
allows to clearly and also easily appreciate recent precision measurements, 
$\lambda{LYB}$ and $\lambda_R$. In Fig. 1 they show some indication in favor of 
new physics, while in 
Figs. 2 and 3 each one is clearly resolved past the 90\% C.L. in favor of
the 
latter. Of course, we still have to wait for a consistent average to be 
determined in the future . Second, our current inability to compute $\lambda$
and $R$ 
with the SM is not a real obstacle to detect new physics in $n\beta d$. Ideally, 
the SM predictions should be a point in the ($\lambda$, R)- plane. Instead, with 
$\lambda$
and $V_{ud}$ remaining as free parameters, the predictions become still 
ideally a line in this plane. Thus, it is still possible to resolve this line from new 
physics contributions. Third, the real limitations for this enterprise come 
from the width of the band around this ideal line. As we have seen, there are 
only two important ingredients in this width. One is the $10^{-4}$ theoretical 
uncertainty in R and the other one is the error bars on $V_{us}$. The first one 
turns out to be not important. It forces us to stop at the band of Fig. 2. 
However, looking at Fig. 3, which assumes that this theoretical error has been 
reduced to zero, we learn that error bars of $V_{us}$ are the real limitation 
for $n\beta d$ to detect new physics. 
In the light of this discussion, we can understand better where future 
efforts should be directed.  On the theoretical side, they should be 
directed towards reducing the theoretical uncertanties in R and in the 
determination of $V_{us}$. On the experimental side, our analysis makes  it 
clear  (see Table 3) that precision measurements must simultaneously  
reduce the error bars on R and $\lambda$. Improving measurements in only one of 
them is useless.
The precision measurements of the other two angular coefficients $\alpha_{e\nu}$
and $\alpha_{\nu}$ lead to the same bands of Figs. 2 and 3. In this sense they 
are no 
longer 
needed. However, if new physics is indeed detected in $n\beta d$, they will be 
very 
useful in delucidating what kind of new physics it would be. Finally, it is 
clear that the determination of $V_{us}$ must be greatly improved, because it 
represents the real limitation for $n\beta d$ to detect new physics.\\
{\bf Acknowledgments.}
The authors are grateful to CONACyT (M\'exico) for partial support.\\

\renewcommand{\baselinestretch}{0.5}

%%%%%%%%%%%%%%%%%%%%%%%%  TABLES 1, 2 and 3 %%%%%%%%%%%%%%%%%%%%%%%%

%%%%%%  TABLE 1  %%%%%%%%%%%%%%%%%%%%%%%%%%%%%%%%%%%%%%%%%%%%%%%%%%% 
\begin{table}
\caption{Ranges for $\lambda$ at fixed values of $R$, in the three cases for the
size of $\sigma_\lambda$ and $\sigma_R$ discussed in the text. For completeness,
we also include the range for $V_{ud}$. $\chi^2_0$ is the minimum of $\chi^2$
and $\lambda_0$ is the corresponding $\lambda$.}
\begin{tabular}{cccccc}
\hline
$R_0$&$\chi^2_0$&$\lambda_0$&
\multicolumn{2}{c}{b o u n d s ~o n~  $\lambda$}&b o u n d s ~o n~  $V_{ud}$\\
&&&$90\%$ C.L.&$95\%$ C.L.&$90\%$ C.L.\\
\hline
1.13209&0.71&1.26988&(1.26581, 1.27464)&(1.26471, 1.27574)&(0.97832, 0.97270)\\
&0.75&1.26988&(1.26911, 1.27072)&(1.26883, 1.27103)&(0.97621, 0.97519)\\
&0.75&1.26990&(1.26952, 1.27041)&(1.26940,1.27060)&(0.97539, 0.97597)\\
&&&&&\\
1.13099&0.65&1.26913&(1.26503, 1.27386)&(1.26393, 1.27496)&(0.97834, 0.97272)\\
&0.71&1.26913&(1.26835, 1.26997)&(1.26807, 1.27029)&(0.97622, 0.97519)\\
&0.69&1.26915&(1.26877, 1.26966)&(1.26865, 1.26985)&(0.97539, 0.97597)\\
&&&&&\\
1.12989&0.61&1.26839&(1.26424, 1.27308)&(1.26314, 1.27418)&(0.97837, 0.97274)\\
&0.67&1.26839&(1.26761, 1.26922)&(1.26734, 1.26953)&(0.97622, 0.97519)\\
&0.64&1.26840&(1.26802, 1.26891)&(1.26790, 1.26910)&(0.97539, 0.97597)\\
&&&&&\\
1.12879&0.56&1.26763&(1.26345, 1.27231)&(1.26235, 1.27341)&(0.97840, 0.97276)\\
&0.64&1.26763&(1.26687, 1.26846)&(1.26659, 1.26876)&(0.97622, 0.97520)\\
&0.59&1.26765&(1.26727, 1.26816)&(1.26715, 1.26835)&(0.97539, 0.97597)\\
&&&&&\\
1.12769&0.54&1.26689&(1.26268, 1.27154)&(1.26158, 1.27264)&(0.97842, 0.97277)\\
&0.60&1.26689&(1.26611, 1.26770)&(1.26582, 1.26801)&(0.97622, 0.97521)\\
&0.56&1.26689&(1.26651, 1.26741)&(1.26639, 1.26760)&(0.97539, 0.97597)\\
&&&&&\\
1.12659&0.52&1.26614&(1.26189, 1.27076)&(1.26079, 1.27186)&(0.97845, 0.97279)\\
&0.58&1.26614&(1.26536, 1.26695)&(1.26508, 1.26725)&(0.97623, 0.97521)\\
&0.52&1.26614&(1.26576, 1.26666)&(1.26564, 1.26685)&(0.97539, 0.97597)\\
&&&&&\\
1.12549&0.50&1.26539&(1.26110, 1.26998)&(1.26001, 1.27108)&(0.97848, 0.97281)\\
&0.56&1.26539&(1.26461, 1.26620)&(1.26433, 1.26652)&(0.97623, 0.97521)\\
&0.50&1.26539&(1.26501, 1.26591)&(1.26489, 1.26610)&(0.97539, 0.97597)\\
\hline
&&&&&\\
&&&&&\\
&&&&&\\
\end{tabular}
\end{table}
%%%%%%%%%%%%%%%%%%%%%%%%%%%%%%%%%%%%%%%%%%%%%%%%%%%%%%%%%%%%%%%%%%%%%%%%%%
%%%%%%%%%%%%%%%%%%%%%  TABLE 2 %%%%%%%%%%%%%%%%%%%%%%%%%%%%%%%%%%%%%%%%%%%%
\begin{table}
\caption{Values of $\chi^2$ at sample points in the ({\large $\lambda$}, $R$) 
plane for current $\sigma_\lambda$ and $\sigma_R$ (upper entries) and for one
tenth of their current values (lower entries).}
\begin{tabular}{c|ccccccccc}
\hline
$R\backslash\lambda$&1.2552&1.2576&1.2600&
1.2624&1.2648&1.2672&1.2696&
1.2720&1.2744\\
\hline
1.13209&33.7&23.7&15.7&9.2&4.7&1.9& 0.8& 1.3&3.4\\
&3352&2353&1440&805&354&87&1.1&32&230\\
&&&&&&&&&\\
1.13099&30.4&21.0&13.4&7.6& 3.6& 1.3& 0.7& 1.7& 4.4\\
&2924&1983&1224&647&253&41&0.8&74&328\\
&&&&&&&&&\\
1.12989&27.3&18.4&11.4& 6.2& 2.7& 0.9& 0.8& 2.3& 5.6\\
&2613&1728&1026&506&168&12&5&133&443\\
&&&&&&&&&\\
1.12879&24.3&16.0& 9.5& 4.9& 1.9& 0.7& 1.0& 3.1& 6.8\\
&2320&1491&845&381&100&1.2&24&208&675\\
&&&&&&&&&\\
1.12769&21.5&13.8& 7.8& 3.8& 1.3& 0.6& 1.5& 4.0& 8.2\\
&2044&1272&682&275&51&0.6&62&302&725\\
&&&&&&&&&\\
1.12659&18.9&11.7& 6.4& 2.8& 0.9& 0.6& 2.0& 5.1&9.9\\
&1784&1069&536&185&17&2.4&117&414&893\\
&&&&&&&&&\\
1.12549&16.4& 9.9& 5.1& 2.0& 0.6& 0.9& 2.8& 6.3&11.7\\
&1543&884&407&114&2.1&18&190&543&1079\\
\hline
\end{tabular}
\end{table}
%%%%%%%%%%%%%%%%%%%%%%%%%%%%%%%%%%%%%%%%%%%%%%%%%%%%%%%%%%
%%%%%%%%%%%%  TABLE 3 %%%%%%%%%%%%%%%%%%%%%%%%%%%%%%%%%%%%%%%%%%

\begin{table}
\caption{Increase of $\chi^2$ as $\sigma_\lambda$ and $\sigma_R$ are reduced as
$1/N$, $N=1,2,...,10$, at the points I$-$IV of figures 2 and 3. The up-downwards
order is also I$-$IV.} 
\begin{tabular}{c|cccccccccc}
\hline
$R\backslash\lambda$&$\sigma_{\lambda}/1$&$\sigma_{\lambda}/2$&$\sigma_{\lambda}/3$
&$\sigma_{\lambda}/4$&$\sigma_{\lambda}/5$&$\sigma_{\lambda}/6$
&$\sigma_{\lambda}/7$&$\sigma_{\lambda}/8$&$\sigma_{\lambda}/9$
&$\sigma_{\lambda}/10$\\
\hline
{\scriptsize $\sigma_{R}/1$}&{\scriptsize 4.8}&{\scriptsize 13.1}&{\scriptsize
21.1}&{\scriptsize 27.3}&{\scriptsize 31.6}&{\scriptsize 34.6}&{\scriptsize
36.7}&{\scriptsize 38.5}&{\scriptsize 39.3}&{\scriptsize 40.1}\\
&{\scriptsize 7.4}&{\scriptsize 20.2}&{\scriptsize 31.5}&{\scriptsize
39.3}&{\scriptsize 44.2}&{\scriptsize 47.6}&{\scriptsize 50.0}&{\scriptsize
52.0}&{\scriptsize 52.9}&{\scriptsize 53.8}\\
&{\scriptsize 0.8}&{\scriptsize 1.0}&{\scriptsize 1.1}&{\scriptsize 1.1}&{\scriptsize 1.2}&{\scriptsize 1.2}&{\scriptsize 1.2}&{\scriptsize 1.2}&{\scriptsize 1.2}&{\scriptsize 1.2}\\
&{\scriptsize 1.2}&{\scriptsize 2.3}&{\scriptsize 3.1}&{\scriptsize 3.8}&{\scriptsize 4.3}&{\scriptsize 4.6}&{\scriptsize 4.8}&{\scriptsize 5.0}&{\scriptsize 5.1}&{\scriptsize 5.2}\\
&&&&&&&&&&\\
{\scriptsize $\sigma_{R}/2$}&{\scriptsize 5.1}&{\scriptsize 16.3}&{\scriptsize 31.9}&{\scriptsize 49.3}&{\scriptsize 66.2}&{\scriptsize 81.5}&{\scriptsize 94.9}&{\scriptsize 106}&{\scriptsize 116}&{\scriptsize 123}\\
&{\scriptsize 8.0}&{\scriptsize 26.6}&{\scriptsize 51.2}&{\scriptsize 77.0}&{\scriptsize 101}&{\scriptsize 121}&{\scriptsize 139}&{\scriptsize 152}&{\scriptsize 164}&{\scriptsize 173}\\
&{\scriptsize 0.8}&{\scriptsize 1.1}&{\scriptsize 1.2}&{\scriptsize 1.3}&{\scriptsize 1.5}&{\scriptsize 1.5}&{\scriptsize 1.6}&{\scriptsize 1.6}&{\scriptsize 1.7}&{\scriptsize 1.7}\\
&{\scriptsize 1.2}&{\scriptsize 2.6}&{\scriptsize 4.2}&{\scriptsize 5.9}&{\scriptsize 7.7}&{\scriptsize 9.4}&{\scriptsize 10.8}&{\scriptsize 12.0}&{\scriptsize 13.1}&{\scriptsize 14.0}\\
&&&&&&&&&&\\
{\scriptsize $\sigma_{R}/3$}&{\scriptsize 5.2}&{\scriptsize 17.1}&{\scriptsize 35.3}&{\scriptsize 58.1}&{\scriptsize 83.4}&{\scriptsize 109}&{\scriptsize 135}&{\scriptsize 160}&{\scriptsize 182}&{\scriptsize 203}\\
&{\scriptsize 8.1}&{\scriptsize 28.1}&{\scriptsize 58.2}&{\scriptsize 94.5}&{\scriptsize 133}&{\scriptsize 172}&{\scriptsize 209}&{\scriptsize 242}&{\scriptsize 272}&{\scriptsize 298}\\
&{\scriptsize 0.8}&{\scriptsize 1.1}&{\scriptsize 1.3}&{\scriptsize 1.5}&{\scriptsize 1.6}&{\scriptsize 1.7}&{\scriptsize 2.0}&{\scriptsize 2.1}&{\scriptsize 2.2}&{\scriptsize 2.3}\\
&{\scriptsize 1.3}&{\scriptsize 2.6}&{\scriptsize 4.5}&{\scriptsize 6.8}&{\scriptsize 9.4}&{\scriptsize 12.1}&{\scriptsize 14.8}&{\scriptsize 17.3}&{\scriptsize 19.8}&{\scriptsize 22.0}\\
&&&&&&&&&&\\
{\scriptsize $\sigma_{R}/4$}&{\scriptsize 5.2}&{\scriptsize 17.5}&{\scriptsize 36.7}&{\scriptsize 62.5}&{\scriptsize 91.8}&{\scriptsize 125}&{\scriptsize 159}&{\scriptsize 194}&{\scriptsize 228}&{\scriptsize 262}\\
&{\scriptsize 8.1}&{\scriptsize 28.8}&{\scriptsize 61.2}&{\scriptsize 103}&{\scriptsize 150}&{\scriptsize 202}&{\scriptsize 254}&{\scriptsize 305}&{\scriptsize 354}&{\scriptsize 400}\\
&{\scriptsize 0.8}&{\scriptsize 1.1}&{\scriptsize 1.3}&{\scriptsize 1.5}&{\scriptsize 1.7}&{\scriptsize 1.9}&{\scriptsize 2.1}&{\scriptsize 2.3}&{\scriptsize 2.5}&{\scriptsize 2.6}\\
&{\scriptsize 1.3}&{\scriptsize  2.7}&{\scriptsize  4.7}&{\scriptsize 7.1}&{\scriptsize 10.1}&{\scriptsize 13.5}&{\scriptsize 17.0}&{\scriptsize 20.6}&{\scriptsize 24.9}&{\scriptsize 27.8}\\
&&&&&&&&&&\\
{\scriptsize $\sigma_{R}/5$}&{\scriptsize 5.2}&{\scriptsize 17.6}&{\scriptsize 37.7}&{\scriptsize 63.9}&{\scriptsize 96.2}&{\scriptsize 133}&{\scriptsize 173}&{\scriptsize 215}&{\scriptsize 259}&{\scriptsize 303}\\
&{\scriptsize 8.1}&{\scriptsize 29.1}&{\scriptsize 62.7}&{\scriptsize 107}&{\scriptsize 160}&{\scriptsize 219}&{\scriptsize 282}&{\scriptsize 347}&{\scriptsize 412}&{\scriptsize 476}\\
&{\scriptsize 0.8}&{\scriptsize 1.1}&{\scriptsize 1.3}&{\scriptsize 1.5}&{\scriptsize 1.8}&{\scriptsize 2.0}&{\scriptsize 2.3}&{\scriptsize 2.6}&{\scriptsize 2.8}&{\scriptsize 3.1}\\
&{\scriptsize 1.3}&{\scriptsize  2.7}&{\scriptsize 4.8}&{\scriptsize 7.5}&{\scriptsize 10.8}&{\scriptsize 14.6}&{\scriptsize 18.3}&{\scriptsize 22.7}&{\scriptsize 27.8}&{\scriptsize 31.7}\\
&&&&&&&&&&\\
{\scriptsize $\sigma_{R}/6$}&{\scriptsize 5.2}&{\scriptsize 17.6}&{\scriptsize 38.0}&{\scriptsize 65.1}&{\scriptsize 98.8}&{\scriptsize 138}&{\scriptsize 182}&{\scriptsize 230}&{\scriptsize 279}&{\scriptsize 330}\\
&{\scriptsize 8.1}&{\scriptsize 29.3}&{\scriptsize 63.5}&{\scriptsize 109}&{\scriptsize 166}&{\scriptsize 230}&{\scriptsize 300}&{\scriptsize 375}&{\scriptsize 452}&{\scriptsize 530}\\
&{\scriptsize 0.8}&{\scriptsize 1.1}&{\scriptsize 1.3}&{\scriptsize 1.5}&{\scriptsize 1.8}&{\scriptsize 2.1}&{\scriptsize 2.4}&{\scriptsize 2.8}&{\scriptsize 3.1}&{\scriptsize 3.4}\\
&{\scriptsize 1.3}&{\scriptsize  2.7}&{\scriptsize 4.7}&{\scriptsize 7.6}&{\scriptsize 10.8}&{\scriptsize 14.7}&{\scriptsize 19.2}&{\scriptsize 24.6}&{\scriptsize 29.8}&{\scriptsize 34.3}\\
&&&&&&&&&&\\
{\scriptsize $\sigma_{R}/7$}&{\scriptsize 5.3}&{\scriptsize 17.7}&{\scriptsize 38.3}&{\scriptsize 65.8}&{\scriptsize 101}&{\scriptsize 141}&{\scriptsize 189}&{\scriptsize 238}&{\scriptsize 293}&{\scriptsize 350}\\
&{\scriptsize 8.1}&{\scriptsize 29.4}&{\scriptsize 64.0}&{\scriptsize 111}&{\scriptsize 169}&{\scriptsize 240}&{\scriptsize 312}&{\scriptsize 394}&{\scriptsize 480}&{\scriptsize 569}\\
&{\scriptsize 0.8}&{\scriptsize 1.1}&{\scriptsize 1.3}&{\scriptsize 1.6}&{\scriptsize 1.9}&{\scriptsize 2.2}&{\scriptsize 2.5}&{\scriptsize 2.9}&{\scriptsize 3.3}&{\scriptsize 3.7}\\
&{\scriptsize 1.3}&{\scriptsize  2.7}&{\scriptsize 7.7}&{\scriptsize 11.2}&{\scriptsize 11.2}&{\scriptsize 15.0}&{\scriptsize 19.7}&{\scriptsize 25.4}&{\scriptsize 30.3}&{\scriptsize 36.1}\\
&&&&&&&&&&\\
{\scriptsize $\sigma_{R}/8$}&{\scriptsize 5.3}&{\scriptsize 17.7}&{\scriptsize 38.4}&{\scriptsize 66.2}&{\scriptsize 102}&{\scriptsize 144}&{\scriptsize 192}&{\scriptsize 245}&{\scriptsize 302}&{\scriptsize 364}\\
&{\scriptsize 8.1}&{\scriptsize 29.4}&{\scriptsize 64.4}&{\scriptsize 112}&{\scriptsize 172}&{\scriptsize 242}&{\scriptsize 321}&{\scriptsize 408}&{\scriptsize 501}&{\scriptsize 598}\\
&{\scriptsize 0.8}&{\scriptsize 1.1}&{\scriptsize 1.3}&{\scriptsize 1.6}&{\scriptsize 1.9}&{\scriptsize 2.2}&{\scriptsize 2.6}&{\scriptsize 3.0}&{\scriptsize 3.4}&{\scriptsize 3.9}\\
&{\scriptsize 1.3}&{\scriptsize  2.7}&{\scriptsize 4.7}&{\scriptsize 7.7}&{\scriptsize 11.3}&{\scriptsize 15.2}&{\scriptsize 20.0}&{\scriptsize 25.4}&{\scriptsize 32.0}&{\scriptsize 37.4}\\
&&&&&&&&&&\\
{\scriptsize $\sigma_{R}/9$}&{\scriptsize 5.3}&{\scriptsize 17.7}&{\scriptsize 38.5}&{\scriptsize 66.6}&{\scriptsize 102}&{\scriptsize 145}&{\scriptsize 194}&{\scriptsize 249}&{\scriptsize 309}&{\scriptsize 374}\\
&{\scriptsize 8.1}&{\scriptsize 29.5}&{\scriptsize 64.6}&{\scriptsize 113}&{\scriptsize 173}&{\scriptsize 245}&{\scriptsize 327}&{\scriptsize 418}&{\scriptsize 516}&{\scriptsize 620}\\
&{\scriptsize 0.8}&{\scriptsize 1.1}&{\scriptsize 1.3}&{\scriptsize 1.6}&{\scriptsize 1.9}&{\scriptsize 2.3}&{\scriptsize 2.7}&{\scriptsize 3.1}&{\scriptsize 3.6}&{\scriptsize 4.0}\\
&{\scriptsize 1.3}&{\scriptsize  2.7}&{\scriptsize 4.7}&{\scriptsize 7.7}&{\scriptsize 11.1}&{\scriptsize 15.4}&{\scriptsize 20.3}&{\scriptsize 25.8}&{\scriptsize 31.9}&{\scriptsize 39.3}\\
&&&&&&&&&&\\
{\scriptsize $\sigma_{R}/10$}&{\scriptsize 5.2}&{\scriptsize 17.8}&{\scriptsize 38.3}&{\scriptsize 66.8}&{\scriptsize 103}&{\scriptsize 146}&{\scriptsize 196}&{\scriptsize 253}&{\scriptsize 315}&{\scriptsize 381}\\
&{\scriptsize 8.1}&{\scriptsize 29.5}&{\scriptsize 64.8}&{\scriptsize 113}&{\scriptsize 175}&{\scriptsize 248}&{\scriptsize 332}&{\scriptsize 425}&{\scriptsize 527}&{\scriptsize 636}\\
&{\scriptsize 0.8}&{\scriptsize 1.1}&{\scriptsize 1.3}&{\scriptsize 1.6}&{\scriptsize 1.9}&{\scriptsize 2.3}&{\scriptsize 2.7}&{\scriptsize 3.2}&{\scriptsize 3.6}&{\scriptsize 4.2}\\
&{\scriptsize 1.3}&{\scriptsize  2.7}&{\scriptsize 4.8}&{\scriptsize 7.7}&{\scriptsize 11.2}&{\scriptsize 15.5}&{\scriptsize 20.5}&{\scriptsize 26.1}&{\scriptsize 33.2}&{\scriptsize 39.1}\\
\hline
\end{tabular}
\end{table}
%%%%%%%%%%%%%%%%%%%%%%%%%%%%%%%%%%%%%%%%%%%%%%%%%%%%%%%%%%%%%%%%%%%%%%%

%%%%%%%%%%%%%%%%%%%%%%%%%%%%%%%  FIGURES  %%%%%%%%%%%%%%%%%%%%%%%%%%%%%%%
\begin{figure}
\epsfxsize=14cm
\centerline{\epsffile{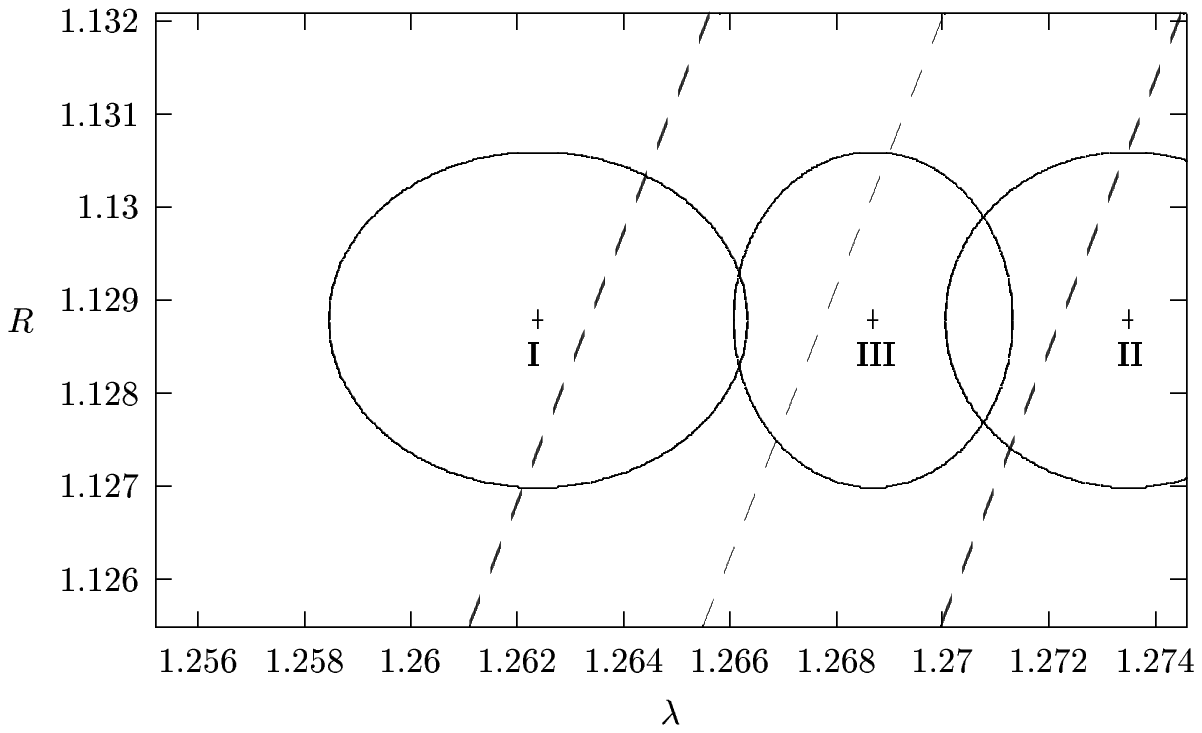}}
\caption{\label{fig:limit}
The SMR, at current values of $\sigma_\lambda$ and $\sigma_R$, is the band within
the dotted lines. See text for other explanations.}
\end{figure}

\begin{figure}
\epsfxsize=14cm
\centerline{\epsffile{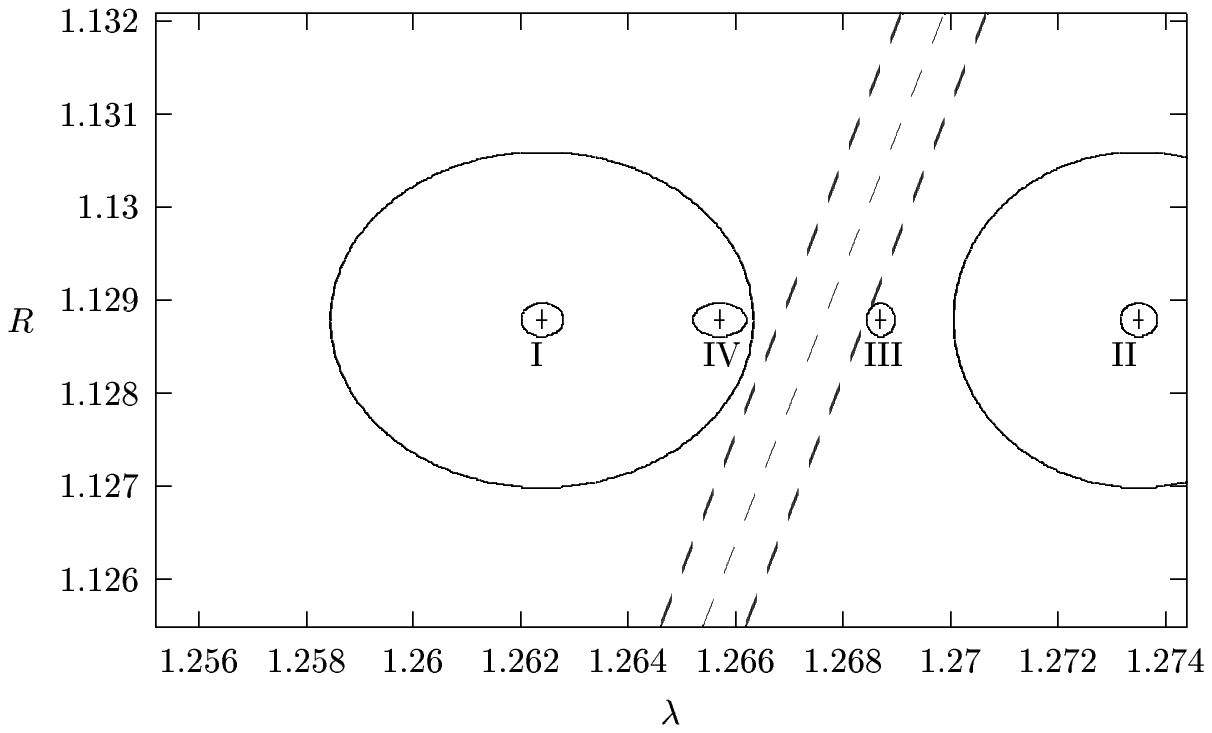}}
\caption{\label{fig:limit}
The SMR, at one tenth of the current values of $\sigma_\lambda$ and $\sigma_R$, 
is the band within the dotted lines. See text for other explanations.}
\end{figure}

\begin{figure}
\epsfxsize=14cm
\centerline{\epsffile{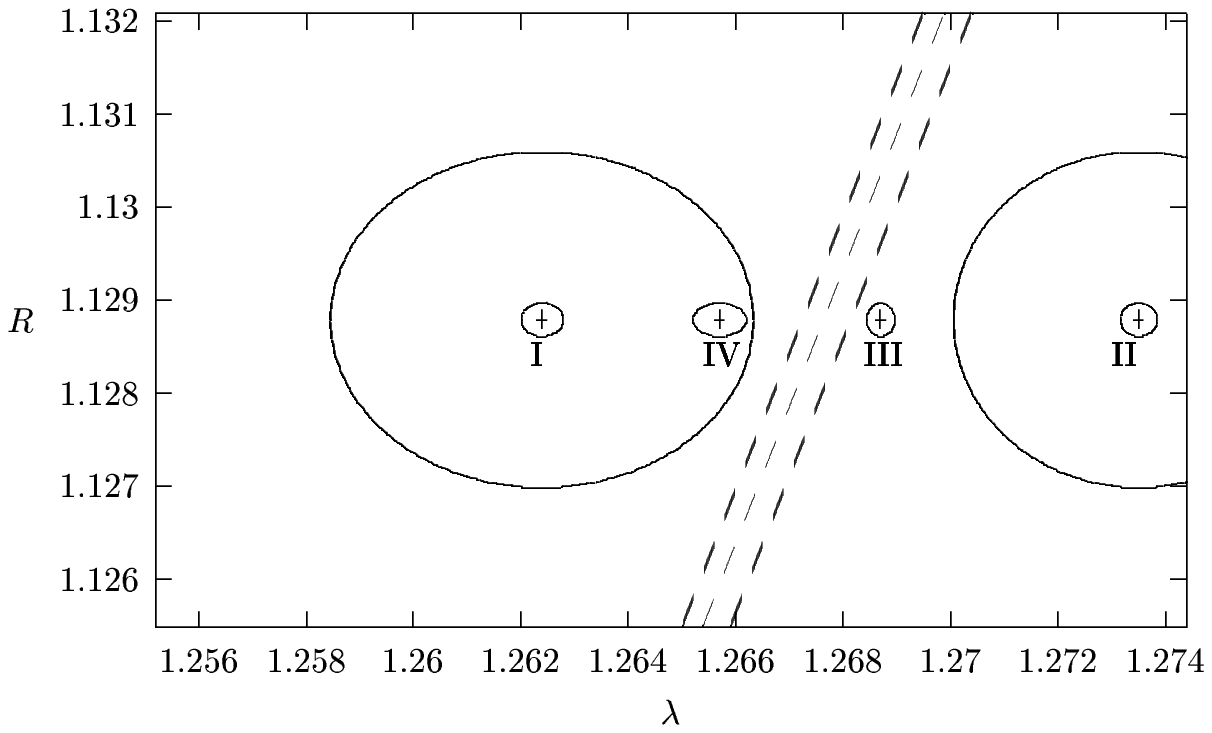}}
\caption{\label{fig:limit}
The SMR, at one thousandth of the current values of $\sigma_\lambda$ and
$\sigma_R$, 
is the band with the dotted lines. See text for other explanations.}
\end{figure}


\newpage

\begin{thebibliography}{99}
\bibitem{abe2} 
H. Abele {\it et al}, Phys. Rev. Lett. {\bf 88} (2002) 211801, and references
therein.
\bibitem{mio} A. Garc\'{\i}a, J.L. Garc\'{\i}a-Luna, G. L\'opez Castro, 
              Phys. Lett. {\bf B500}, 66 (2001).
\bibitem{3}
D.H. Wilkinson, Nucl. Instrum. Methods {\bf A335},(1993) 172, 182, 201  and 
references therein.
\bibitem{pdg}
D. E. Groom, {\it et al}, Particle Data Group, Eur. Phys. J. {\bf C15}, 
(2000) 1.
\bibitem{liaud} P. Liaud {\it et al}, Nucl. Phys. {\bf A612}, (1997) 53.
\bibitem{bopp} P. Bopp {\it et al}, Phys. Rev. Lett. {\bf 56}, (1986) 919.
\bibitem{yero} B. Yerozolimsky {\it et al}, Phys. Lett. {\bf B412}, (1997) 240.
\bibitem{rei} J. Reich {\it et al}, {\it V Int. Seminar on Interaction
of Neutrons with Nuclei}, Dubna 1997.
\bibitem{2} D.H. Wilkinson, Z. Phys. {\bf A348}, (1994) 129.
\end{thebibliography}
\end{document}